\documentclass[12pt]{article}
\usepackage{amsfonts}
\usepackage{latexsym,amsmath}
\usepackage{graphicx}
\usepackage{placeins}
\usepackage{slashbox}
\usepackage{amssymb,array}
\makeatletter \oddsidemargin 0in \evensidemargin 0in \textwidth 16cm
\newcommand{\es}{\end{subsection}}

\begin{document}
 \pagestyle{empty}


\title{\large \bf Cryptanalysis and Improvement of Jiang et al.'s Smart Card Based Remote User Authentication Scheme}

\author{\small Dheerendra Mishra\thanks{E-mail:~{dheerendra@maths.iitkgp.ernet.in} }, Ankita Chaturvedi and Sourav Mukhopadhyay\\
\small Department of Mathematics,\\ 
\small Indian Institute of Technology Kharagpur,\\
\small  Kharagpur 721302, India\\}

\date{}
 \maketitle

\begin{abstract}

Smart card based remote user password authentication schemes are one of the user-friendly and scalable mechanism to establish secure communication between remote entities. These schemes try to ensure secure and authorized communication between remote entities over the insecure public network. Although, most of the existing schemes do not satisfy desirable attributes, such that resistance against attacks, user anonymity and efficiency. In 2012, Chen et al. proposed a robust smart cased based remote user authentication scheme to erase the weaknesses of Sood et al.'s scheme. Recently, Jiang et al. showed that Chen et al.'s scheme is vulnerable to password guessing attack. Furthermore, Jiang et al. presented a solution to overcome the shortcoming of Chen et al.'s scheme. In the paper, we show that Jiang et al.'s scheme  is still vulnerable to insider attack, on-line and off-line password guessing attack and user impersonation attack. Their scheme also fails to ensure perfect forward secrecy and user's anonymity. Moreover, It does not provide efficient login and user-friendly password change phase. Further, to overcome these drawbacks, we present a modify scheme which reduces the computation overhead and satisfies all desirable security attributes where Jiang et al.'s scheme failed.
\end{abstract}
\textbf{keywords:} {Smart card; Password based authentication; Cryptanalysis; Anonymity.}

\section{Introduction}{\label{intro}}

The advancements in technology have made the Internet an efficient and scalable tool to utilize for various online services. However, an adversary may have full control over the network and can perform various kinds of attacks. Therefore, to ensure authorized and secure communication,  user and server should mutually authenticate each other and draw a session key. The smart card based authentication protocols are designed and developed to ensure secure and authorized communication between remote user and server~\cite{khan2011cryptanalysis}.

 In 2009, Xu et al.~\cite{xu2009improved} presented an improved smart card based password authentication scheme to overcome the weaknesses of Lee et al.'s scheme~\cite{lee2005improved}. Xu et al. also claimed that their scheme satisfies all the desirable security attributes.  Although, in 2010, Sood et al.~\cite{sood2010improvement} showed that Xu et al.'s scheme is vulnerable to offline password-guessing attack and forgery attacks. They also presented an improvement of Xu et al.'s scheme. In the same year, Song~\cite{song2010advanced} also demonstrated that an adversary can retrieve the  stored information from the smart card and can perform user impersonation attack. Further, he presented an enhanced authentication scheme using smart card to overcome the weaknesses of Xu et al.'s scheme. In 2012, Chen et al.~\cite{chen2012robust} pointed out that the improvements presented by both Song and Sood et al. are still vulnerable to known attacks. Chen et al. showed that Sood et al.'s scheme does not achieve mutual authentication as it supports only one way authentication where only server verifies the user's authenticity. In addition, they identified the inefficiency of Sood et al.'s scheme in the detection of incorrect input. Chen et al.  also demonstrated the offline password guessing attack on the Song's scheme. Moreover, they proposed an efficient authentication scheme. Recently, Jiang et al.~\cite{jiang2013improvement} analyzed Chen et al.'s scheme and showed that Chen et al.'s scheme does not resist password guessing attack. They also proposed a solution to erase the drawback of password guessing attack. Unfortunately, Jiang et al.'s scheme does not erase password guessing attack efficiently and it is still vulnerable to off-line and on-line password guessing attack.  It does not resist insider attack and user impersonation attack. Additionally, it does not support session key verification which helps to enhance data security and integrity.

 An adversary can eavesdrop the user and the server interaction as they communicate via public channel. Therefore, secrecy of a consumer's identity should be supported during message exchange. Otherwise, it may give an opportunity to the adversary to collect the users specific information that enables him to track the consumer's current location and login history. Unfortunately, none of the  aforementioned password based authentication schemes~\cite{xu2009improved,lee2005improved,lee2005improvement,sood2010improvement,song2010advanced,chen2012robust,jiang2013improvement} protect anonymity. Additionally, a user should allow  to recover his lost smart card. Although these schemes do not present smart card revocation phase where an authorized user can recover his lost smart card with the help of  server.

In this article, we present a brief review of Jiang et al.'s scheme and demonstrate the vulnerability of their scheme to off-line and on-line password guessing attack, insider attack and user impersonation attack. We point out inefficiency of Jiang et al.'s scheme to protect user anonymity and to present user-friendly password change phase and inefficient login phase. Further, we proposed an enhanced password based authenticated key agreement scheme using smart card to overcome the weaknesses of Jiang et al.'s scheme.

The rest of the paper is organized as follows: Section~\ref{review} presents the brief review of Jiang et al.'s scheme.  Section~\ref{crypt} points out the weakness of Jiang et al.'s scheme. Finally, conclusion is drawn in Section \ref{conclusion}.

\renewcommand{\labelitemi}{$\bullet$}
\section{Review of Jiang et al.'s Scheme}\label{review}

In 2013, Jiang et al.~\cite{jiang2013improvement}  proposed an improvement of Chen et al.'s~\cite{chen2012robust} remote user's authentication scheme. Their scheme, registration and password change phases are similar to Chen et al.'s scheme. However the login \& authentication phase are different to overcome the weaknesses of Chen et al.'s scheme. This schemes has the following four phases:\\
\begin{enumerate}
  \item Registration phase
  \item Login phase
  \item Authentication phase
  \item Password change phase
\end{enumerate}


In the beginning of the system, the server chooses two large prime numbers $p$ and $q$ such that $p = 2q + 1$. It also selects the master secret key $x\in Z_q$ and a one way hash function $h(\cdot) : \{0,1\}^{*}\rightarrow Z^{*}_p$. Then, the registration, login and authentication phases execute as follows:
 \subsection{Registration Phase}
 To achieve a valid smart card, a user proceeds as follows:
\begin{description}

\item [ Step 1.] $U_i$ chooses a unique identity $ID_i$ and password $PW_i$. Then, he submits $ID_i$ and  $PW_i$ to $S$ via a secure channel.

\item[ Step 2.] $S$ computes $B_i= h(ID_i)^{(x+ PW_i)} \pmod p$.

    \item [\bf Step 3.] $S$ embeds the parameters $\{B_i, h(\cdot), p, q\}$ into the smart card and issues to $U_i$. It also stores $ID_i$ in its ID table.

\end{description}

\subsection{Login  Phase}

\begin{description}

\item [\bf Step 1.] $U_i$ enters his smart card into the card reader and inputs $ID_i$ and $PW_i$.

\item[\bf Step 2.]  The smart card selects a random number $\alpha \in Z_q^{*}$ and calculates the following values at time $T_i$:\\
 \begin{eqnarray*}
   C_i &=& B_i/ h(ID_i)^{PW_i} \pmod p \\
   D_i &=& h(ID_i)^{\alpha} \pmod p \\
  W_i &=& (C_i)^{\alpha} \pmod p \\
  M_i &=& h(ID_i\parallel C_i\parallel D_i\parallel W_i \parallel T_i)
 \end{eqnarray*}

 Then, it sends the  message $\{ID_i, D_i, M_i, T_i\}$ to $S$.

\end{description}


\subsection{Authentication  Phase}
\begin{description}

\item [\bf Step 1.] When $S$ receives the message at time $T_i^{'}$, it  verifies the existence of $ID_i$ in its database. If $ID_i$ exists, then verifies $T_U^{'}-T_U\leq \triangle T$, where $\triangle T$ is the valid time delay in message transmission. If conditions does not hold, it terminates the session. Otherwise, it computes the following values:

    \begin{eqnarray*}
      C'_i &=& h(ID_i)^{x} \pmod p \\
     W_i' &=& (D_i')^{x} \pmod p \\
  M_i' &=& h(ID_i\parallel C_i'\parallel D_i\parallel W_i' \parallel T_i)
    \end{eqnarray*}

    \item [\bf Step 2.] $S$ verifies $M_i^{'} =? ~ M_i$. If verification does not hold, it rejects the request. Otherwise, $U_i$ is authenticated by $S$.

\item [\bf Step 3.] $S$ takes the current timestamp $T_S$ and computes $M_S = h(ID_i\parallel W_i'\parallel T_S)$, then transmits the message $\{ID_i, M_S, T_S\}$ to $U_i$.

\item [\bf Step 4.] Upon receiving the message at time $T_S^{'}$, $U_i$ validates  $T_S^{'}- T_S\leq \triangle T$. If verification succeeds, $U_i$ verifies $M_S =? ~h(ID_i\parallel W_i\parallel T_S)$.

\item [\bf Step 5.] $U_i$ and $S$ computes their respective session keys $SK = h(W_i) = h(W_i')$.
\end{description}


\subsection{Password Change Phase}
A user can change his password as follows:

\begin{description}

\item [\bf Step 1.] $U_i$ enters the smart card into a card reader, then inputs identity $ID_i$, old password $PW_i$ and new password $PW_{new}$.

\item[\bf Step 2.]  The smart card interacts with the server $S$ to confirm the correctness of old password $PW_i$ by executing login and authentication phase. If old password verification holds, the smart card  computes

   $ B_{new}= B_i\cdot h(ID_i)^{PW_{new}}/h(ID_i)^{PW_i} \pmod p.$

 \item [\bf Step 3.] Finally, the smart card replaces $B_i$ with $B^{new}_U$.

\end{description}


\renewcommand{\labelitemi}{$-$}

\section{Cryptanalysis of Jiang et al.'s Scheme}\label{crypt}

In this section, we will discuss the flaws of Jiang et al's scheme. After analysis, we find that their scheme cannot resist some of the known attacks such as insider, password guessing attack and user impersonation attack. \nocite{eisenbarth2008power,kocher1999differential} 
%
%

\renewcommand{\labelitemii}{$-$} 
 \renewcommand{\labelitemi}{$\bullet$}

\subsection{User anonymity}
The leakage of the user's specific information enables the adversary to track the user's current location and login history~\cite{juang1999anonymous}. Although user's anonymity ensures user's privacy by preventing an attacker from acquiring user's sensitive personal information. Moreover, anonymity makes remote user authentication mechanism more robust as an attacker could not track which users are interacting with the server.

The straightforward way to preserve anonymity is to conceal user's real identity during communication. However, Jiang et al.'s scheme takes user's real identity in login message. It shows that Jiang et al.'s scheme does not protect anonymity.

\subsection{Insider Attack:}

In general, a user uses the same password for several accounts because it is difficult to remember several distinct passwords for different accounts. When a user submits his password in its original form to the server, a malicious insider can know the user's password. This gives the opportunity to a malicious insider to access user's accounts which are protected with the same passwords. Unfortunately, Jiang et al.'s scheme does not prevent insider attack as user submits its original password to the server.


\subsection{On-line password guessing attack}

In Jiang et al.'s scheme, the server does not track the login requests, that is, server does not count the unsuccessful login request. It provides an opportunity to an adversary to perform online password guessing attack as server does not deny incorrect repeated login request. An adversary can successfully perform on-line password guessing attack as follows:

\begin{description}

\item[ Step $1$.] Adversary could achieve stored secret information $<B_i, h(\cdot), p, q>$ from the lost smart card. Moreover, he can intercept the user login message $\{ID_i ,D_i ,M_i ,T_i\}$ and achieve user's identity $ID_i$.

 \item[Step $2$.] The adversary guesses the password $PW_i^*$ and selects a value $e$, then computes the following values:

 \begin{eqnarray*}
   C_i^* &=& B_i/ h(ID_i)^{PW_i^*} \pmod p \\
   D_E &=& h(ID_i)^{e} \pmod p \\
  W_E &=& (C_i^*)^{e} \pmod p \\
  M_E &=& h(ID_i\parallel C_i^*\parallel D_E\parallel W_i^* \parallel T_E)
 \end{eqnarray*}

   Then, the adversary sends the message $\{ID_i ,D_E ,M_E ,T_E\}$ to $S$.

    \item[ Step $3$.]  The verification of $ID_i$ and $T_E$ holds, as $ID_i$ is user's identity and $T_E$ is fresh timestamp used by adversary.
   Then, the server computes the following values:

          \begin{eqnarray*}
      C_i &=& h(ID_i)^{x} \pmod p \\
     W_i^* &=& (D_E)^{x} \pmod p
    \end{eqnarray*}
Then, it verifies  $M_E =? h(ID_i\parallel C_i'\parallel D_E\parallel W_E' \parallel T_i)$.   If verification does not hold, it rejects the request. Otherwise, responds with a valid message.

\item[ Step $4$.] If verification fails at server's side, adversary repeats {\bf Step} 2 and  {\bf Step} 3. Otherwise, the password guessing attack will be succeeded.

\end{description}

%
%
%
%
%
%
%

\subsection{Off-line password guessing attack}
An adversary can guess a legitimate user's password with the help of retrieve value $B_i$ and $ID_i$ from the stolen smart card $SC\{B_i, h(\cdot), p, q\}$ using power analysis attack~\cite{eisenbarth2008power,kocher1999differential} and intercepted login message $<ID_i, a_i, r_i>$, respectively. An adversary can guess the password as follows:

 \begin{description}
 \item[Step 1.] An adversary intercepts the user's login message $<ID_i, D_i, M_i, T_i>$ and retrieves user's identity $ID_i$.

 \item[Step 2.] In Jiang et al.'s scheme, the server does not verify the registration of identity, that is, whether the identity submitted for registration is already registered or not. It provides opportunity to an adversary to achieve user's secret key $h(ID_i)^x$ using user's identity $ID_i$ as follows:

\begin{itemize}
  \item $E$ selects a random value $PW_E$, then submits $ID_i$ and $PW_E$ to $S$.
  \item  Upon receiving the request, $S$ computes  $B_E= h(ID_i)^{(x+ PW_E)}\pmod p$.

  \item $S$ embeds the parameters $\{B_E, h(\cdot), p, q\}$ into the smart card and provides it to $E$. 

 \item The adversary extracts $B_E$ from the smart card and computes user's secret key as follows:

    $$h(ID_i)^x = B_E/h(ID_i)^{PW_E}\pmod p$$

 \end{itemize}

  \item[\bf Step 3.] An attacker guesses the value $PW_i^*$ and computes $X_i^* = B_i\oplus h(ID_i)^{PW_i^*}$, then verifies  $X_i^* =?~ h(ID_i)^x$.

 \item [\bf Step 4.] If the verification succeeds, considers $PW_i^*$ as the user's password. Otherwise, he repeats {\bf Step 3}.
  \end{description}

\subsection{User impersonation attack}

An adversary can masquerade as a legitimate user by successfully login to the server as follows:


\begin{itemize}

  \item An adversary intercepts user's login message $\{ID_i, D_i, M_i, T_i\}$ and retrieves user's identity $ID_i$ from it. 

   \item The adversary achieves user's secret key $h(ID_i)^x$ using user's identity $ID_i$ as discussed in off-line password guessing attack.

  \item The adversary  chooses a random number $e \in Z_q^{*}$ and computes the following values:

      \begin{eqnarray*}
   D_E &=& h(ID_i)^{e} \pmod p \\
  W_E &=& (h(ID_i)^x)^{e} \pmod p \\
  M_E &=& h(ID_i\parallel h(ID_i)^x\parallel D_E\parallel W_E \parallel T_E)
 \end{eqnarray*}

 Then, he sends the  message $\{ID_i, D_E, M_E, T_E\}$ to $S$ where $T_E$ is the current timestamp.

\item When $S$ receives the message at time $T_E'$, it  verifies the $ID_i$ and  $T_U^{'}-T_U\leq \triangle T$. Both the conditions hold as adversary uses registered user's identity and current timestamp. Then, $S$ computes $C_i = h(ID_i)^{x} \pmod p$ and $W_E' = (D_E)^{x} \pmod p$, and verifies  \begin{eqnarray*}
  M_E &=?& h(ID_i\parallel C_i\parallel D_E\parallel W_E'\parallel T_E).
    \end{eqnarray*}

The verification holds as $W_E' = (D_E)^{x} \pmod p = h(ID_i)^{ex} \pmod p = W_E$.

\item Since, the  verification holds, $S$ authorized the message and computes $M_S = h(ID_i\parallel W_E'\parallel T_S)$ where $T_S$ is the current timestamp. It sends the message $\{ID_i, M_S, T_S\}$ to $U_i$. $S$ also computes the session key $SK = h(W_E')$.

\item $E$ intercepts the message $\{ID_i, M_S, T_S\}$ and calculates the session key  $SK = h(W_E)$.
\end{itemize}

The discussion shows that an adversary can successfully login to the server and compute the session key.

\subsection{Time synchronization problem}

To identify the replay attack, smart card based authentication schemes use timestamp mechanism $(T'_U -T_U \leq \Delta T)$, where $T_U$ is the time when the message is sent, $T'_U$ is the message receiving time and  $\Delta T$ is the predetermined time delay in message transmission. In general, a user device clock (local clock) may not synchronize with the server. So, if the interval of time delay in message transmission $\Delta T$ is too small, the server may not identify the valid message and deny legitimate request as it does not satisfy the condition. Further, if the interval of the time delay in message transmission $\Delta T$ is too large, the server may not identify the replay attack.

In jiang et al.'s scheme timestamp is used to resist replay attack.  In general, all hardware clocks are imperfect,  local clock of user device  may drift away from the server  in time~\cite{sivrikaya2004time}. Therefore,  the observed time or durations of the valid time intervals may differ for each device in the network. If the clock on user device which is used for time stamping, is differ by a significant amount, the valid login message  $\{ID_i, D_i, M_i, T_i\}$ does not pass the condition $T_U^{'}-T_U\leq \triangle T$. It shows that a valid may fail to login to the server due to time synchronization problem.



\subsubsection{Perfect forward secrecy}
In Jiang et al.'s scheme an adversary can compute the session key  using compromised master key $x$ of the server  as follows:

\begin{itemize}
  \item  To compute the session key $SK = h(W_i)$, an adversary has to compute $W_i = (D_i)^{x} \pmod p$.

  \item  An adversary can achieve $D_i = h(ID_i)^{\alpha} \pmod p $ from $\{ID_i, D_i, M_i, T_i\}$ as the adversary can achieve old transmitted messages via public channel.

\item An adversary can compute $W_i = (D_i)^{x} \pmod p$ using compromised master key $x$.
\end{itemize}

Since, the compromise of master key may result compromise of session key, it shows that proposed scheme does not ensure perfect forward secrecy.


\subsection{Inefficient login phase}

The smart card cannot verify the input in Jiang et al.'s scheme and executes the login session in case of incorrect input. It shows the inefficiency of scheme in incorrect input detection. It causes extra computation and communication overhead. If a user may input incorrect password or identity due to mistake, then following cases arises:

\noindent\textbf{Case 1:} If a user inputs wrong password $PW_i^*$ due to mistake.

\begin{itemize}

\item  The smart card selects a random number $\alpha \in Z_q^{*}$ and calculates the following values at time $T_i$:\\
 \begin{eqnarray*}
   C_i^* &=& B_i/ h(ID_i)^{PW_i^*} \pmod p \neq h(ID_i)^{x} \pmod p~ as~ PW_i \neq PW_i^*\\
   D_i &=& h(ID_i)^{\alpha} \pmod p \\
  W_i^* &=& (C_i^*)^{\alpha} \pmod p \\
  M_i^* &=& h(ID_i\parallel C_i^*\parallel D_i\parallel W_i^* \parallel T_i)
 \end{eqnarray*}

 Then, it sends the  message $\{ID_i, D_i, M_i, T_i\}$ to $S$.

\item  When $S$ receives the message at time $T_i^{'}$, it  verifies the $ID_i$ existence in its database and  $T_U^{'}-T_U\leq \triangle T$. The verification holds as identity $ID_i$ is correct and smart  card uses current timestamp, then it computes the following values:

    \begin{eqnarray*}
      C_i &=& h(ID_i)^{x} \pmod p \\
     W_i &=& (D_i)^{x} \pmod p \\
  M_i &=& h(ID_i\parallel C_i\parallel D_i\parallel W_i \parallel T_i)
    \end{eqnarray*}

    \item When $S$ verifies $M_i =? ~ M_i^*$. The verification does not hold as $W_i \neq W_i^*$, then server rejects the request.
\end{itemize}

\noindent\textbf{Case 2:} If a user inputs incorrect identity $ID_i^*$.

\begin{itemize}

\item  The smart card selects a random number $\alpha \in Z_q^{*}$ and calculates the following values at time $T_i$:\\
 \begin{eqnarray*}
   {C_i'}^* &=& B_i/ h(ID_i^*)^{PW_i} \pmod p \neq h(ID_i)^{x} \pmod p ~as~ ID_i \neq ID_i^*\\
   D_i^* &=& h(ID_i^*)^{\alpha} \pmod p \\
  {W_i'}^* &=& ({C_i'}^*)^{\alpha} \pmod p \\
  {M_i'}^* &=& h(ID_i^*\parallel {C_i'}^*\parallel D_i^*\parallel {W_i'}^* \parallel T_i)
 \end{eqnarray*}

 Then, it sends the  message $\{ID_i^*, D_i^*, {M_i'}^*, T_i\}$ to $S$.

\item  When $S$ receives the message at time $T_i^{'}$, it  verifies the existence of $ID_i^*$ in its database. The verification does not hold as identity $ID_i^*$ is incorrect.
\end{itemize}

\subsection{Unfriendly password change phase}
To change the password of the smart card, a user has to establish an authorized session with the server, that means, a user cannot change his password freely. This shows the inefficiency of Jiang et al.'s scheme.


\renewcommand{\labelitemii}{$-$}
 \renewcommand{\labelitemi}{$\bullet$}

\renewcommand{\labelitemi}{$ $}
\section{Proposed scheme}\label{proposed}
Jiang et al. tried to overcome the weaknesses of Chen et al.'s scheme by modifying its login and authentication phase. Although, they failed to satisfy desirable security attributes. To overcome the weaknesses of Jiang et al.'s scheme, we propose an improved scheme which comprises the following phases:

\begin{itemize}
  \item (i)~ Initialization
  \item (ii)~ Registration
  \item (iii)~Login
  \item (iv)~Authenticated key agreement
  \item (v)~Password change
  \item (v)~ Smart card revocation
\end{itemize}

\subsection{Initialization}

In the beginning, server chooses two large prime numbers $p$ and $q$ such that $p = 2q + 1$. It selects a secret key $x \in Z_q$, say, master key. It also chooses a one way hash function $h(\cdot) : \{0, 1\} \rightarrow Z_p^*$, for example SHA-1.

\subsection{Registration Phase}
First, a non registered user submits his registration request along with identity and password to the server. The user does not submit password in its original form, he submits hashed output of the password to prevent insider attack. Upon receiving the user's request, the server verifies identity registration, that means, identity is already registered or not. If identity is already registered with some other user, it asks for new identity. Otherwise, it completes user's registration and provides a smart card with personalized parameters to the user. The detailed description of the procedure is as follows:

\begin{description}

\item [ Step 1.] $U$ selects a random number $a$ and chooses an identity $ID_i$ and  a password $PW_i$ of his choice. He computes $W = h(PW_i\oplus a)$ and submits ($ID_i, W$) with new user registration request to $S$ via secure channel.

\item[ Step 2.] Upon receiving the $U$'s registration request, $S$ verifies  credential of identity $ID_i$. If server found $ID_i$ in its database, that means, $ID_i$ is registered with some other user, the server asks for the new identity. Otherwise, it computes $X_i =  h(ID_i||N||ID_{SC}||x)$ and $B = X_i\oplus W$ where $ID_{SC}$ is the issued smart card secret identity, $NID$ is a generated pseudonym identity and $N = 0$ if $U$ is a new user, otherwise $N = N + 1$.

 \item [\bf Step 3.] $S$ personalizes the mart card by embedding the parameters $\{NID, B, h(\cdot), p, q\}$ into the smart card. Then, it provides the smart card $SC\{NID, B, h(\cdot), p, q\}$ to $U$ via secure channel. Additionally, $S$ maintains a database of registered users, say, users' record table. The entry $N||ID_{SC}||ID_i$ is added corresponding to $NID$ into users' record table. It also stores $N$ corresponding to $ID_i$ in registered user's database.


    \item [\bf Step 4.] Upon receiving the smart card,  $U$ computes $L = a\oplus h(ID_i\oplus PW_i)$ and $V = h(ID_i||a||PW_i)$. He stores $L$ and $V$ into the smart card. Finally, the smart card stores the parameters $\{NID, B, L, V, h(\cdot), p, q\}$.

   \end{description}
\begin{figure}[h]
  \includegraphics[width=16cm]{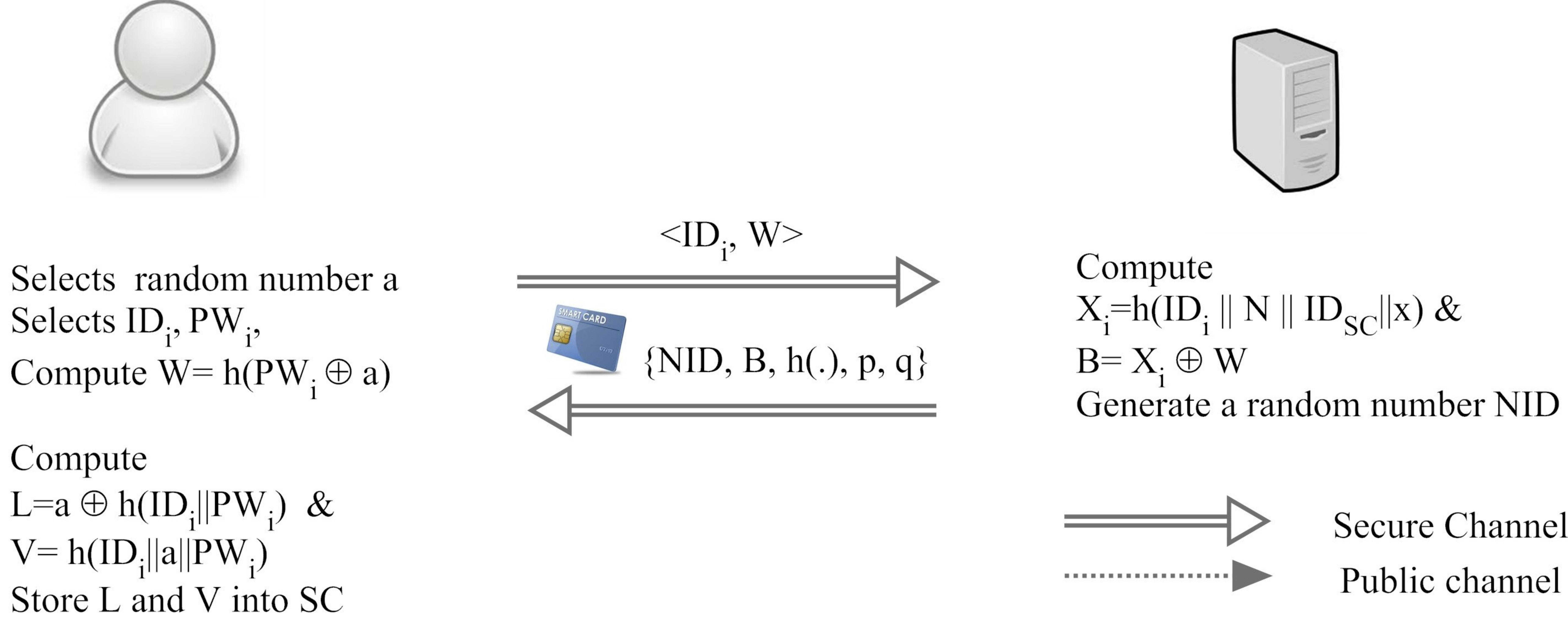}\\
  \caption{The pictorial representation of registration phase}\label{pr-1}
\end{figure}

\subsection{Login Phase}
To established authorized session with the server, user sends a login message to the server.  To generate the login message, user initiates login message by inputting his identity and password to the smart card. First, the smart card verifies the correctness of input parameters. If verification fails, it terminates the session. Otherwise, it executes the login session which works as follows:

\begin{description}

\item [Step 1.] Compute $a = L\oplus h(ID_i\oplus PW_i)$ then verify $V =?~h(ID_i||a||PW_i)$. If verification does not hold, terminate the session. Otherwise, goto {\em Step 2.}

\item[Step 2.] Compute $W = h(PW_i||a)$ and then $X_i = B\oplus W$.

\item[Step 3.] Select a random number $\alpha\in Z_q^*$ and compute $D_i = h(ID_i)^{\alpha}~\mbox{mod}~p$ and $M_1 = h(ID_i||D_i||X_i)$.

\item[Step 4.] Send the login message $<NID, D_i, M_1>$ to $S$.

\end{description}


\begin{figure}[h]
  \includegraphics[width=16cm]{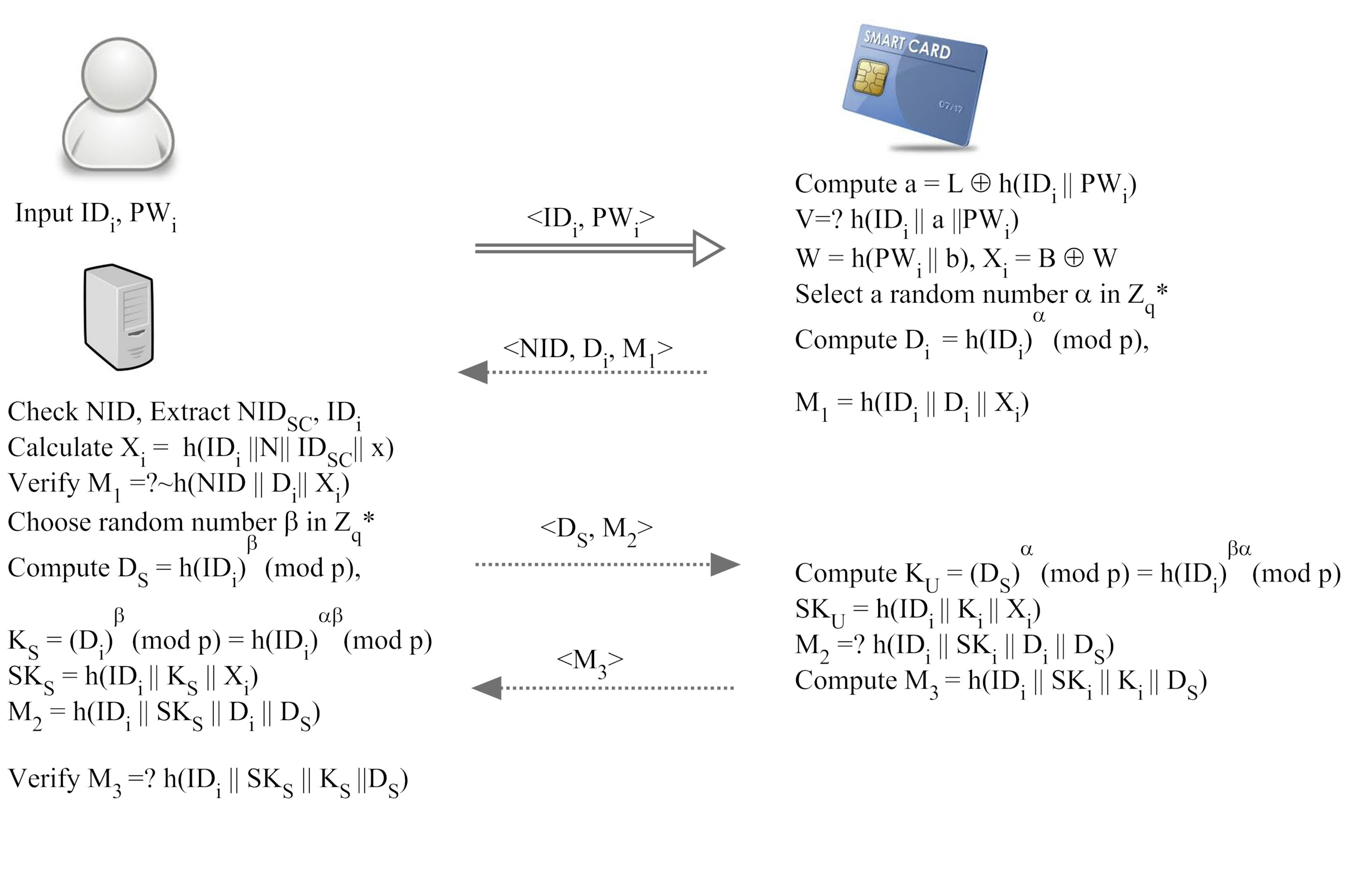}\\
  \caption{The pictorial representation of login and authentication phase}\label{pr-1}
\end{figure}

\subsection{Authenticated key agreement phase}

Upon receiving user's login request, the server verifies the authenticity of the message. If verification succeeds, it responses with a valid message. Moreover, user also verify the authenticity of the server. On the success of mutual authenticity, user and server compute the session key and verify it. The detailed description of mutual authentication and session key establishment is as follows:


\begin{description}

\item [\bf Step 1. ] Upon receiving the message $<NID, D_i, M_1>$, $S$ checks the value $NID$ in users' record table. If $NID$ does not exist, it denies the request. Otherwise, it extracts the values $N, ID_{SC}$ and $ID_i$ corresponding to $NID$ from its database. It calculates $X_i =  h(ID_i||N||ID_{SC}||x)$,  then verifies $M_1 =?~h(ID_i||D_i||X_i)$. If verification does not hold, it denies the login request. Otherwise, $S$ chooses a random number $\beta\in Z_q^*$ and computes $D_S = h(ID_i)^{\beta}~\mbox{mod}~p$, $K_{S} = (D_i)^{\beta}~\mbox{mod}~p = h(ID_i)^{\alpha\beta}~\mbox{mod}~p$ and the session key $SK_{S} = h(ID_i||K_{S}||X_i)$.

\item [\bf Step 2. ] $S$  computes $M_2 = h(ID_i||SK_{S}||D_i||D_S)$ and sends the response message $<D_S, M_2>$ to $U$.

\item[\bf Step 3.] Upon receiving the message $<D_S, M_2>$, $U$ computes $K_{i} = (D_S)^{\alpha}~\mbox{mod}~p = h(ID_i)^{\beta\alpha}~\mbox{mod}~p$ and the session key $SK_{i} = h(ID_i||K_{i}||X_i)$ and then verifies $M_2 =?~h(ID_i||SK_{i}||D_i||D_S)$. If verification does not hold, the session is terminated. Otherwise, the server is authenticated  and session key is verified. 

\item[\bf Step 4.] $U$ computes $M_3 = h(ID_i||SK_{i}||K_{i}||D_S)$ and sends $<M_3>$ to $S$.

\item[\bf Step 5.] Upon receiving the message $<M_3>$, $S$  verifies $M_3 =?~h(ID_i||SK_{S}||K_S||D_S)$. If verification does not hold, the session is terminated. Otherwise, $U$ is authenticated  and session key is verified.
\end{description}


\subsection{Password Change Phase}
The proposed scheme presents user-friendly password change phase where a user with correct identity and password can change the password without server assistance. The proposed password change phase, first verifies the correctness of input parameters (identity and password). If verification does not succeed, it terminate the session. Otherwise, it executes the password change phase. The description of password change phase is as follows:

\begin{description}

\item [\bf Step 1. ] $U$ inserts his smart card into the card reader and inputs identity $ID_i$, old password $PW_i$ and a new password $PW_{new}$.

\item [\bf Step 2. ] The smart card computes $a = L\oplus h(ID_i\oplus PW_i)$ and then verifies  $V =?~h(ID_i||a||PW_i)$. If verification does not succeed, it terminates the  session. Otherwise, run {\em Step P3.}

 \item [\bf Step 3. ] The smart card computes $W = h(PW_i||a)$ and $W_{new} = h(PW_{new}||a)$ then $B_{new} =  B_i\oplus W\oplus W_{new}$, $L_{new} = a\oplus h(ID_i\oplus PW_{new})$ and $V_{new} = h(ID_i||a||PW_{new})$. Then, it replaces $B$ with $B_{new}$, $L$ with $L_{new}$ and $V$ with $V_{new}$.

\end{description}
\begin{figure}[h]
  \includegraphics[width=16cm]{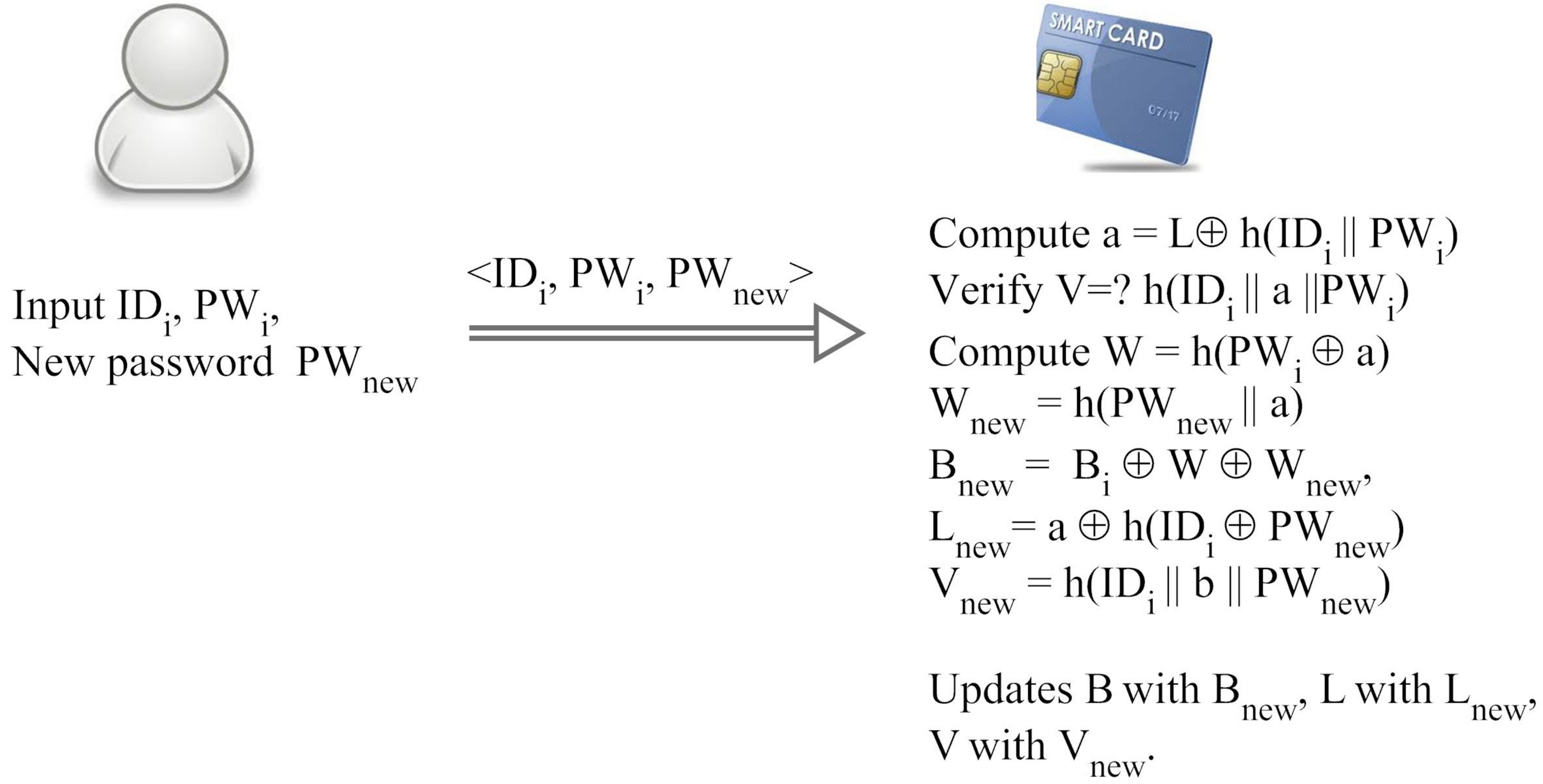}\\
  \caption{The pictorial representation of password change phase}\label{pr-1}
\end{figure}
\subsection{Smart card revocation}

If a legal user lost his smart card, then user can get a new smart card from the server as follows:
\begin{description}

\item [\bf Step 1. ] $U$ chooses a password $PW_i'$ and a random number $a'$ then computes  $W' = h(PW_i'||a')$. He submits his new smart card request with $ID_i$ and $W'$ to $S$ via secure channel.

\item [\bf Step 2. ] Upon receiving the request, $S$ verifies the registration of $U$. If $U$'s identity $ID_i$ does not exist in server's registered user's list, it terminates the session. Otherwise, it achieves $N$ corresponding to $ID_i$.

\item[ Step 3.] $S$ takes $ N = N+1$ and select $ID_{SC}'$ and $NID'$ then computes $X_i' =  h(ID_i||N||ID_{SC}'||x)$ and  $B' = X_i' + W'$.

\item[ Step 4.] $S$ personalizes $U$'s smart card by embedding the security parameters $\{NID', B', h(\cdot), p, q\}$ into the smart card and provides it to $U$ via secure channel. $S$ also updates $N$ with $N+1$ and adds the entry $(N||ID_{SC}'||ID_i)$  corresponding to $NID'$ into users' record table. 

\item [ Step 5.] Upon receiving the smart card, $U$ performs the {\em Step $4$} of registration phase.

\end{description}

\begin{figure}[h]
  \includegraphics[width=16cm]{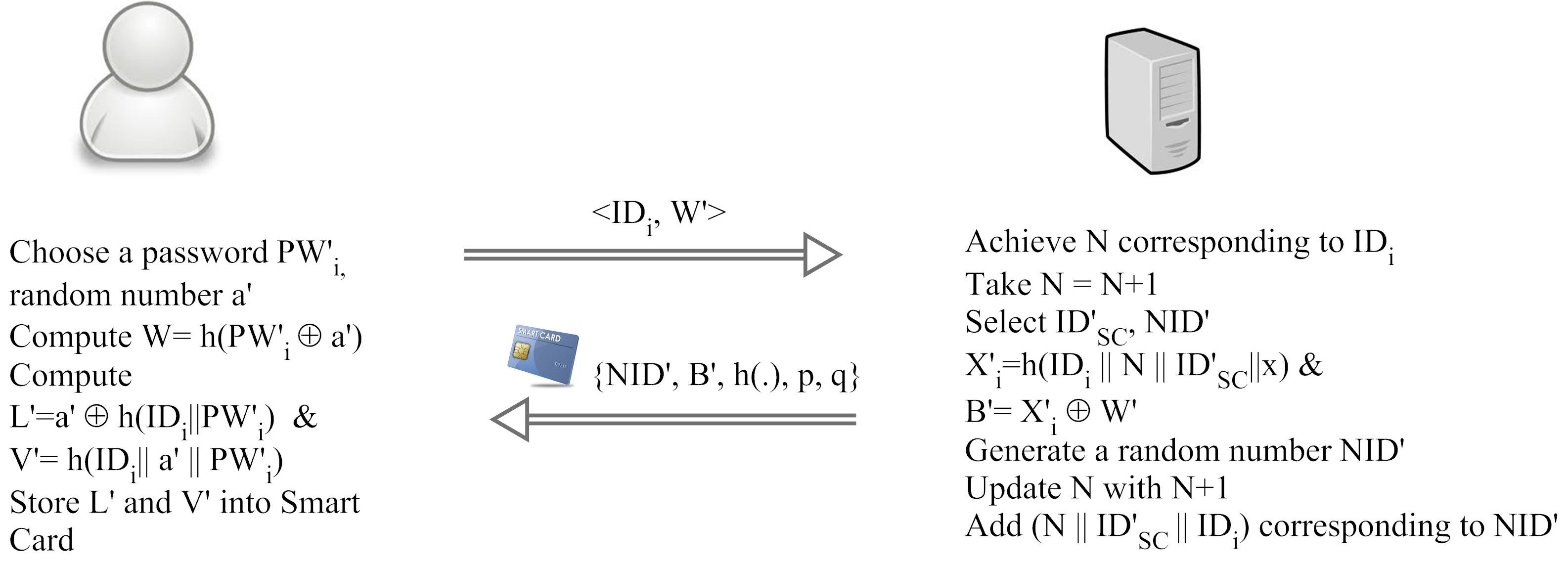}\\
  \caption{The pictorial representation of revocation phase}\label{pr-1}
\end{figure}
\renewcommand{\labelitemii}{$-$}
 \renewcommand{\labelitemi}{$\bullet$}
 
\section{Analysis}

\subsection{Security analysis}\label{security}

The detailed security analysis of the proposed scheme to verify `how the scheme satisfying the security requirements' is as follows:\\

\subsubsection{User anonymity}
  The login message  and smart card keeps dynamic identity $NID$ which is a random value. So, no information can be collected about $ID_i$ using $NID$. Moreover, an adversary may try to guess $ID_i$ using the conditions $V = h(ID_i||a||PW_i)$ or $M_1 = h(ID_i||D_i||X_i)$. However, the identity guessing cannot succeeded because of the following facts:

\begin{itemize}
  \item To verify the guessed identity $ID_i^*$ with $V = h(ID_i||a||PW_i)$, $a$ is needed. Although to compute $a$ from $L = a\oplus h(ID_i\oplus PW_i)$, the password $PW_i$ is needed. The password is only known to the user.

  \item To verify the guessed identity $ID_i^*$ with $M_1 = h(ID_i||D_i||X_i)$, the user's secret key $X_i$ is needed. To extract $ID_i$ from $B = X_i\oplus W$, it requires user's password $PW_i$ as $W = h(PW_i\oplus a)$.
\end{itemize}

\subsubsection{Insider attack}
The user submits $W$ to the server instead of $PW_i$, where $W = h(PW_i\oplus a)$. Therefore, an insider cannot achieve consumer password as hash function is one way. Moreover, an adversary cannot guess the password using  $W = h(PW_i\oplus a)$ as user does not submit random value $a$ to the server.
\subsubsection{ Stolen smart card attack}
 Let the lost or stolen smart card of a user is achieved by an adversary. The adversary can retrieve the parameters $\{NID, B, L, V\}$ from the smart card and may try to use this information to login to the server. However, this attempt cannot be succeeded in the proposed scheme which is justified as follows:

  \begin{itemize}

    \item To generate a valid login message $<NID, D_i, M_1>$, an adversary has to compute $M_1 = h(ID_i||D_i||X_i)$.

    \item To compute $M_1$, the user's secret key $X_i$ and identity $ID_i$ are needed.

    \item Neither the smart card nor the transmitted messages includes $ID_i$. To compute $X_i$ from $B$, password is needed. Therefore, an adversary cannot achieve $ID_i$ and $X_i$.

  \end{itemize}

 Since the password is only known to the user and the identity is secret, an adversary cannot generate a  valid login message using stolen smart card. This shows that the proposed scheme withstands stolen smart card attack.
\subsubsection{Off-line password guessing attack}

An adversary may try to guess user's password. To guess the password, he can retrieve the information $NID$, $B = X_i\oplus W$, $L = a\oplus h(ID_i\oplus PW_i)$ and $V = h(ID_i||a||PW_i)$ from the smart card using power analysis attack~\cite{messerges2002examining}. Then, an adversary may try to guess password as follows:

\begin{itemize}
  \item The adversary guesses the password $PW^*$.
  \item To verify the guessed password $PW^*$ with the condition $V =?~h(ID_i||a||PW_i^*)$, requires user's identity $ID_i$.

  \item Neither the smart card stores $ID_i$ nor any transmitted message.

  \item To guess the password, an adversary has to guess user's identity $ID_i$.

  \item   If  $ID_i$ is of $n$ characters, then the probability to guess a correct $n$ characters of $ID_i$ is approximately $1/2^{ 6n}$.

  \item If the password is of $m$ characters, then the probability to guess of $ID_i$ and $PW_i$ at the same time is approximately $1/2^{ 6n + 6m}$.

\end{itemize}

The above discussion shows that an adversary has to guess both identity and password at the same time. Since, it is computationally infeasible to guess both value at the same time as  probability approximated to $1/2^{6n + 6m}$, the proposed scheme resists password guessing attack.

\subsubsection{On-line password guessing attack}
 An active adversary may try to verify guessed password by generating valid login message. To generate valid login message, an adversary may use the retrieved information $\{NID, B, L, V\}$ from the smart card, and intercepted previously transmitted login messages $<NID, D_i, M_1>$. However, the adversary cannot create a valid login message to verify the guessed password. It is justified from the following discussion:

 \begin{itemize}
  \item Let the adversary guess the password $PW^*_i$.

 \item To verify the guessed password $PW^*$, an adversary tries to generate a valid login message $<NID, D_i^*, M_1^*>$, where $D_i^* = h(ID_i)^{e}~\mbox{mod}~p$ and $M_1^* = h(ID_i||D_i^*||X_i^*)$ for a random value $e$. It is equivalent to achieve $X_i$ from $B = X_i\oplus W$ using guessed password, where $W = h(PW_i\oplus a)$ and $ a = L\oplus h(ID_i\oplus PW_i)$.

 \item To compute $a$, the $ID_i$ is needed along with $PW_i$ as $ a = L\oplus h(ID_i\oplus PW_i)$.


 \item To compute $M_1^* = h(ID_i||D_i^*||X_i)$, user's identity $ID_i$ is also needed.

 \item Neither the smart card nor the transmitted messages include  $ID_i$. Therefore, an adversary has to guess identity along with password at the same time.

\item To perform on-line password guessing attack, an adversary has to guess both identity and password at the same time. As we already discussed that it is infeasible.

\end{itemize}

It is clear from the discussion that an adversary cannot successfully perform on-line password guessing attack. 

 \subsubsection{Replay attack}

An adversary can eavesdrop user's communication, and  can intercept and record old transmitted messages  $<NID, D_i, M_1>$, $<D_S, M_2>$ and $<M_3>$. Then, he can try to replay the old login message.

 \begin{itemize}
   \item Let  adversary replay the message $<NID, D_i, M_1>$.
   \item Upon receiving the message $<NID, D_i, M_1>$, $S$ checks the value $NID$ in users' record table and finds it as adversary repeats user's valid message. It extracts the values $N, ID_{SC}, ID_i$ corresponding to $NID$ from its database. It computes $X_i =  h(ID_i||N||ID_{SC}||x)$ and verifies $M_1 =?~h(ID_i||D_i||X_i)$. The verification succeeds.

   \item $S$ chooses a random number $\beta'\in Z_q^*$ and computes $D_S' = h(ID_i)^{\beta'}~\mbox{mod}~p$, $K_{S}' = (D_i)^{\beta'}~\mbox{mod}~p$,  $SK_{S}' = h(ID_i||K_{S}'||X_i)$ and $M_2' = h(ID_i||SK_{S}'||D_i||D_S')$, then sends the message $<D_S', M_2'>$ to $U$.

\item The adversary intercepts the message $<D_S', M_2'>$ and try to respond.

\item If adversary respond with the old transmitted message $<M_3>$, where $M_3 = h(ID_i||SK_{i}||K_{i}||D_S)$ and $K_{i} = (D_S)^{\alpha}~\mbox{mod}~p = h(ID_i)^{\beta\alpha}~\mbox{mod}~p$. The server identify the replay attack as $\beta' \neq \beta$

  \item An adversary may also try to respond with $<M_3'>$. To compute $M_3'$, an adversary has to compute $K_{i}' = h(ID_i)^{\beta'\alpha}~\mbox{mod}~p$ as $M_3' = h(ID_i||SK_{i}'||K_{i}'||D_S')$. 
  
 \item To compute $K_{i}' = h(ID_i)^{\beta'\alpha}~\mbox{mod}~p$ from $D_i = h(ID_i)^{\alpha}~\mbox{mod}~p$ and $D_S' = h(ID_i)^{\beta'}~\mbox{mod}~p$ is equivalent to Computational Diffie–Hellman (CDH) problem which is hard.


\item Since the adversary cannot respond with the valid message, the server terminates the session.

 \end{itemize}
\subsubsection{User impersonation attack}
An adversary can masquerade as a legitimate user by successfully login to the server. However, the proposed scheme can resist this attack as follows:
\begin{itemize}
  \item An adversary may try to login to the server using replay attack. Although the proposed scheme resist replay attack.

    \item An adversary mat try to generate a valid login message $<NID, D_i', M_1'>$ for a random value $e$, where $D_i' = h(ID_i)^{e}~\mbox{mod}~p$ and $M_1 = h(ID_i||D_i'||X_i)$. However, an adversary cannot compute $D_i$ and $M_1$ correctly as he cannot  achieve $ID_i$ and $X_i$. It is justified as follows:

     \begin{itemize}

    \item To compute $M_1$,  $X_i$ and $ID_i$ are needed as $M_1 = h(ID_i||D_i'||X_i)$.

    \item Neither the smart card nor the transmitted messages includes $ID_i$. So, an adversary cannot achieve $ID_i$.

    \item To compute $X_i$ from $B$, the password is needed. Since the password is only known to the user, an adversary cannot achieve $X_i$.

  \end{itemize}

\end{itemize}

This shows that the proposed scheme resists user impersonation attack.

\subsubsection{Server impersonation attack}
An adversary can masquerade as a server and try to respond with valid message to the user as follows:

\begin{itemize}

  \item When an user sends a login message $<NID, D_i', M_1'>$ to the server, the adversary intercept the message, where $D_i' = h(ID_i)^{\alpha'}~\mbox{mod}~p$ and $M_1' = h(ID_i||D_i'||X_i)$.

\item An adversary may try to respond using old message of server $<D_S, M_2>$, where $M_2 = h(ID_i||SK_{S}||D_i||D_S)$, $D_i = h(ID_i)^{\alpha}~\mbox{mod}~p$, $D_S = h(ID_i)^{\beta}~\mbox{mod}~p$,  $K_{S} = h(ID_i)^{\alpha\beta}~\mbox{mod}~p$ and $SK_{S} = h(ID_i||K_{S}||X_i)$. However, the user can identity the replay of old message as follows:

    \begin{itemize}

    \item Upon receiving the message $<D_S, M_2>$, the user computes $K_{i}' = (D_S)^{\alpha'}~\mbox{mod}~p = h(ID_i)^{\beta\alpha'}~\mbox{mod}~p$ and the session key $SK_{i}' = h(ID_i||K_{i}'||X_i)$

        \item The user verifies $M_2 =?~h(ID_i||SK_{U}'||D_i'||D_S)$. The verification does not hold as $\alpha' \neq \alpha$ and so  $K_{i}' \neq  K_{S}$ and $ D_i' \neq D_i$.

      \end{itemize}

 \item An adversary may try to generate the valid login message $<D_E, M_2^*>$ for a random value $e$, where $D_E = h(ID_i)^{e}~\mbox{mod}~p$, $M_2^* = h(ID_i||SK_{E}||D_i'||D_E)$ and $SK_{E} = h(ID_i||K_{E}||X_i)$. However, an adversary cannot compute $M_2^*$ correctly due to the following facts:

     \begin{itemize}

     \item To compute $M_2^* = h(ID_i||SK_{E}||D_i'||D_E)$, an adversary has to compute $SK_{E} = h(ID_i||K_{E}||X_i)$.

    \item To compute $SK_{E}$,  $X_i$ and $ID_i$ are needed.

    \item Neither the smart card nor the transmitted messages includes $ID_i$. So, an adversary cannot achieve $ID_i$.

    \item To compute $X_i$ from $B$, the password is needed. Since the password is only known to the user, an adversary cannot achieve $X_i$.

  \end{itemize}

\end{itemize}

This shows that the proposed scheme resists server impersonation attack.

\subsubsection{Time synchronization problem}
Deploying the timestamp method to resist the replay attack, requires the cost of implementing clock synchronization, that is, the clock time of the all the registered users and the server must not fluctuate out of a small range. To overcome this problem, the proposed scheme uses random number instead of timestamp to verify the freshness of message. 

\subsubsection{Mutual authentication}
The server verifies the authenticity of user with the condition $M_1 =?~h(ID_i||D_i||X_i)$. Since to compute $M_1$, user's identity $ID_i$ and secret key $X_i$ is needed, therefore, the server can correctly verify the user's authenticity as adversary cannot achieve  $ID_i$ and $X_i$. The user verifies the authenticity of user with the condition $M_2 =?~ h(ID_i||SK_{S}||D_i||D_S)$, where $SK_{S} = h(ID_i||K_{S}||X_i)$. Since no unauthorized party can compute $SK_{S} = h(ID_i||K_{S}||X_i)$ as it requires  $ID_i$ and $X_i$. So, the user can correctly verify the server authenticity.

\subsubsection{Session Key agreement}
The user and the server compute  the session keys $SK_{i} = h(ID_i||K_{i}||X_i)$ and $SK_{S} = h(ID_i||K_{S}||X_i)$, respectively. The computed session keys  $SK_{i}$ and $SK_{S}$ are same at both ends as
\begin{eqnarray*}
  K_{S} &=&(D_i)^{\beta}~\mbox{mod}~p \\
    &=& h(ID_i)^{\alpha\beta}~\mbox{mod}~p \\
    &=& h(ID_i)^{\beta\alpha}~\mbox{mod}~p\\
    &=&(D_S)^{\alpha}~\mbox{mod}~p \\
   &=&   K_{i}
\end{eqnarray*}


\subsubsection{Session key verification}
The user verifies whether the server has computed the session key correctly using the condition $M_2 = h(ID_i||SK_{S}||D_i||D_S)$. The server  verifies whether the user has computed the session key correctly using the condition $M_3 = h(ID_i||SK_{i}||K_{i}||D_S)$. Since, both $M_2$ and $M_3$ include the session key, the user and the server can correctly verify the established session key.

\subsubsection{Key freshness}

 Each session key $SK_{S} = h(ID_i||K_{S}||X_i)$, where $K_{S} = h(ID_i)^{\alpha\beta}~\mbox{mod}~p$, involves random numbers $\alpha$ and $\beta$. The random values $\alpha$ and $\beta$ are fresh for each session. Uniqueness of these values for each session, guaranties the unique key for each session. The unique key construction for each session ensures the key freshness property.

 \subsubsection{ known key secrecy}
  If the previously established session key $SK_{S} = SK_{i} =  h(ID_i||K_{S}||X_i)$ is compromised, then the compromised session key reveals no information about other session keys due to following reasons:

 \begin{itemize}
   \item Each key is hashed with one way hash function, therefore, no information can be retrieve from the session key.
   \item Each session key involves random numbers which guarantees different key for each session.
 \end{itemize}
Since no information about other established session keys from the compromised session key is extracted. This shows that proposed scheme achieves known key secrecy. \\


\subsubsection{Forward secrecy}
Forward secrecy states that compromise of user long-term secret key does not become the reason to compromise of established session keys.
In proposed scheme, if the user long-term secret key $X_i$ is compromised, then an adversary cannot compute the session key as he cannot achieve $ID_i$ and cannot compute $h(ID_i)^{\alpha\beta} ~\mbox{mod}~p$ which is justified as follows:

\begin{itemize}
  \item Neither the smart card nor the transmitted messages include $ID_i$, therefore, an adversary can not achieve $ID_i$.
  \item  To  compute $h(ID_i)^{\alpha\beta} ~\mbox{mod}~p$ from  $D_i = (ID_i)^{\alpha} ~\mbox{mod}~p$ and $(ID_i)^{\beta} ~\mbox{mod}~p$ is equivalent to computational Diffie-Hellman (CDH) problem. Since CDH problem is hard, therefore, no unauthorized user can compute  $K_{S}$ or $K_{i}$ using $D_i$ and $D_S$.
\end{itemize}

\subsubsection{Perfect forward secrecy}
In perfect forward secrecy scenario, an adversary cannot compute the session key with the compromised master key of the server. Although if the server master key $x$ is compromised, an adversary may compute the user's secret key $X_i =  h(ID_i||N||ID_{SC}||x)$, but he cannot compute the session key which is justified as follows:

\begin{itemize}
  \item  To compute the session key $h(ID_i||K_{S}||X_i)$, an adversary has to compute $K_S = h(ID_i)^{\alpha\beta} ~\mbox{mod}~p$.
  \item  To  compute $h(ID_i)^{\alpha\beta} ~\mbox{mod}~p$ using  $D_i = (ID_i)^{\alpha} ~\mbox{mod}~p$ and $(ID_i)^{\beta} ~\mbox{mod}~p$ is equivalent to computational Diffie-Hellman (CDH) problem. Since CDH problem is hard, therefore, an adversary cannot compute  $h(ID_i)^{\alpha\beta} ~\mbox{mod}~p$ using $D_i$ and $D_S$.
\end{itemize}

Since, the compromised of master key does not mean compromised of session key, it shows that proposed scheme ensures perfect forward secrecy.

\subsubsection{Known session-specific temporary information attack}
 If the short-term keys or temporary secrets, say, $\beta$ and $\alpha$ are compromised, then an attacker may try to construct the session key $SK_{S} = SK_{i} = h(ID_i||K_{S}||X_i)$ using $\beta$ and $\alpha$. Because, to compute the session key user's identity $ID_i$ and $X_i$ are needed along with $\beta$ or $\alpha$. Since neither smart card stores $ID_i$ nor transmitted messages associate it.  Additionally, the user secret key $X_i$ is protected with password. An adversary cannot achieve $ID_i$ and $X_i$. This shows that the proposed scheme resists Known session-specific temporary information attack.


\subsubsection{Efficient login  phase}

In the proposed scheme, smart cards can correctly identify the incorrect input as follows:\\

\noindent{\em Case-1.} If the smart card receives incorrect password $PW_i^*$ instead of  $PW_i$ then
\begin{itemize}
  \item  The smart card retrieves $a^* = L\oplus h(ID_i\oplus PW_i^*)$ and verifies $V_i =? h(ID_i||a^*||PW_i^*)$.

  \item The verification does not hold as $V =  h(ID_i||a||PW_i)$ and $PW_i \neq PW_i^*$.
\end{itemize}

\noindent{\em Case-2.} If the smart card receives incorrect identity $ID_i^*$ then
\begin{itemize}
  \item  The smart card retrieves $a'^* = L\oplus h(ID_i^*\oplus PW_i)$ and verifies $V_i =? h(ID_i^*||a'^*||PW_i)$.

  \item The verification does not hold as $V =  h(ID_i||a||PW_i)$ and $ID_i^* \neq ID_i$.
\end{itemize}

\noindent{\em Case-3.} If the smart card receives incorrect identity $ID_i^*$ and password $PW_i^*$ then
\begin{itemize}
  \item  The smart card retrieves $a''^* = L\oplus h(ID_i^*\oplus PW_i^*)$ and verifies $V_i =? h(ID_i^*||a''^*||PW_i^*)$.

  \item The verification does not hold as $V =  h(ID_i||a||PW_i)$, $ID_i^* \neq ID_i$ and $PW_i \neq PW_i^*$.
\end{itemize}

In all the above cases the smart card can detect the incorrect input. This shows that proposed scheme has efficient login phase.

\subsubsection{User-friendly and efficient password changes phase}
The user is allowed to change his password without server assistance. This makes proposed scheme user-friendly. Moreover, the smart card verifies the correctness of identity and password using the condition $V =?~  h(ID_i||a||PW_i)$. If the verification does not succeed, the smart card terminates the session. Otherwise, it allows to change the password. Since the smart card can verify the correctness of input efficiently, a user can change his password correctly without any mistake.\\


 The comparison of proposed scheme with Xu et al.'s, Song's, Sood et al.'s, Chen et al.'s and Li et al.s' schemes is presented in Table-{1}. If the scheme prevent attack or satisfies the attribute, the symbol $(\surd)$ is used. otherwise, the symbol { ($\times$)} is used.

 \FloatBarrier
\begin{table}
 {\centering
  \caption{Comparison of the proposed scheme with related schemes for different  desirable security attributes}\label{t2}

 \begin{tabular}{lccccccc} \hline

   \backslashbox{ Security attributes }{Schemes}

&\cite{xu2009improved}&\cite{song2010advanced}&\cite{sood2010improvement} & \cite{chen2012robust}  & ~\cite{jiang2013improvement}  & Proposed \\ \hline

 User anonymity       &  $\times$ & $\times$  &  $\times$   & $\times$          &   $\times$   &    $\surd$\\

 Insider Attack   &  $\times$ & $\times$  &  $\times$   & $\times$        &   $\times$    &    $\surd$  \\

 On-line password guessing attack    & $\times$  &  $\times$   & $\times$  &  $\times$ &  $\times$ &  $\surd$\\
 Off-line password guessing attack     & $\times$  &  $\times$   & $\times$  &  $\times$ &  $\times$ &  $\surd$\\

Forward secrecy    & $\surd$ & $\times$ & $\surd$ & $\surd$    &    $\surd$     &  $\surd$    \\

 Known session keys  attack  & $\surd$ &  $\surd$ &  $\surd$ &  $\surd$   &     $\surd$ &  $\surd$ \\

 User impersonation attack &  $\times$ & $\times$  &  $\times$   & $\times$       &    $\times$    & $\surd$\\

Server impersonation attack  & $\surd$  & $\times$ &  $\surd$ & $\times$ &    $\surd$       & $\surd$ \\



Replay attack   & $\surd$ &  $\surd$ &  $\surd$ &  $\surd$   &     $\surd$ &  $\surd$ \\

Time synchronization problem  &  $\times$ & $\times$  &  $\times$   & $\times$          &   $\times$   &    $\surd$\\


Mutual authentication  & $\surd$ & $\surd$ &  $\times$   & $\surd$  &  $\surd$ &  $\surd$\\


Efficient login phase &  $\times$ &  $\times$ &  $\times$ & $\times$  & $\surd$ &    $\surd$  \\

Efficient password change  phase   & $\surd$ &  $\surd$ &  $\surd$ &  $\surd$   &     $\surd$ &  $\surd$ \\

User-friendly password  change  phase  &  $\times$ &  $\times$ &  $\times$ & $\times$  & $\times$ &    $\surd$  \\

Session key agreement  & $\surd$   &  $\surd$ &  $\surd$    &   $\surd$ &   $\surd$    &    $\surd$\\
Session key verification &  $\times$ & $\times$  &  $\times$   & $\times$          &   $\times$   &    $\surd$\\
Smart card revocation &  $\times$ & $\times$  &  $\times$   & $\times$          &   $\times$   &    $\surd$\\

           \hline
\end{tabular}
}

\vspace{.2cm}

 \end{table}



\subsection{Performance analysis}\label{performance}

In general, the smart cards have limited storage space and computation capacity. Therefore, the authentication protocol must give priority to the efficiency due to resource constraints in smart card~\cite{liao2009secure}.  In this section, we show the efficiency analysis of proposed schemes with similar password based remote user authentication  protocols based on smart card. Let $T_h$, $T_E$, $T_M$, $T_S$ and $T_X$ denote the time complexity of hash function, exponential operation, multiplication/division operation, symmetric encryption/decryption operation and XOR operation, respectively. It is well known that the time complexity of XOR operation is negligible as compared to two other operations. So, we do not take $T_X$ into account. In general, the time complexity associated with $T_h$, $T_E$ and $T_X$ can be more or less expressed as $T_E >> T_h >> T_X$~\cite{potlapally2006study,wong2001performance}.

\begin{table}
 {
  \caption{Computation cost comparison of proposed scheme with related schemes}\label{t2}

  \begin{tabular}{|l|c|c|c|c|} \hline
{Schemes}
    &Registration & Login  &  Authentication & Password change \\ \hline

Xu et al.  & $T_E + 2T_h$ & $3T_h + 2T_E$       & $6T_h  + 2T_E$  & $7T_h + 4T_E$\\

Song &$T_E + 2T_h$  & $2T_h + 1T_S$             & $6T_h + 1T_S + 1T_E$  & $8T_h + 2T_S + 1T_E$\\

Sood et al. &$2T_E + T_h$ & $2T_h + 2T_M + 3T_E$   & $4T_h + 1T_M + 2T_E$ & $4T_h + 5T_M + 7T_E$\\

Chen et al. & $T_h + T_E$ &         $2T_h + 2T_M + 2T_E$      & $6T_h + 1T_M + 1T_E$ & $6T_h + 5T_M + 5T_E$\\

Jiang et al. & $T_h + T_E$  & $2T_h + T_M + 3T_E$     & $6T_h  + 2T_E$  & $6T_h + 3T_M + 7T_E$\\

Proposed    & $4T_h$ &  $4T_h + T_E$   &   $8T_h + 3T_E$   &   $6T_h$\\

    \hline
  \end{tabular}

}

\end{table}


\section{Conclusion}\label{conclusion}

The presented article analyzes Jiang et al.'s scheme and demonstrates the weakness of their schemes. This investigation shows that their scheme is vulnerable to on-line and off-line password guessing attack, insider attack and user impersonation attack. It also fails to protect anonymity and to present efficient login and use-friendly password change phase. Further, we have presented an improved smart card based anonymous user authentication scheme to remove all the drawbacks of Jiang et al.'s scheme. Moreover, the proposed scheme present smart card revocation phase where a user can achieve lost smart card with the help of server without registering again. 

%
\bibliographystyle{dheerendra}
\bibliography{jiang-biblo}

\end{document}